\documentclass{PoS}

\title{Measuring the growth of galaxy clusters}

\ShortTitle{Measuring the growth of galaxy clusters}

\author{\speaker{Antonaldo Diaferio}\\
        Università di Torino, Dipartimento di Fisica, via P. Giuria 1, I-10125, Torino, Italy\\
        Istituto Nazionale di Fisica Nucleare (INFN), Sezione di Torino, Torino, Italy\\
        E-mail: \email{diaferio@ph.unito.it}}

\abstract{We suggest how we can use  the mass profile of galaxy clusters beyond their virial radius to 
measure their mass accretion rate,  a key prediction of structure formation models.
The mass profile can be estimated by applying the caustic technique to dense redshift surveys 
of clusters and their outskirts, where dynamical equilibrium does not necessarily hold. 
An additional probe of the mass growth of clusters is their mass fraction 
in substructures. We show that the caustic technique, that identifies cluster substructures
as a by-product, returns catalogs of substructures  with mass larger than a  few $10^{13}h^{-1}M_\odot$ 
that are  between  60\% and 80\% complete, depending on the density of the redshift survey.}

\FullConference{Frontiers of Fundamental Physics 14\\
                 15-18 July 2014\\
                 Aix Marseille University (AMU) Saint-Charles Campus, Marseille, France}

\begin{document}

\section{The mass accretion rate of galaxy clusters}

\begin{figure}
 \centerline{\includegraphics[width=0.7\textwidth]{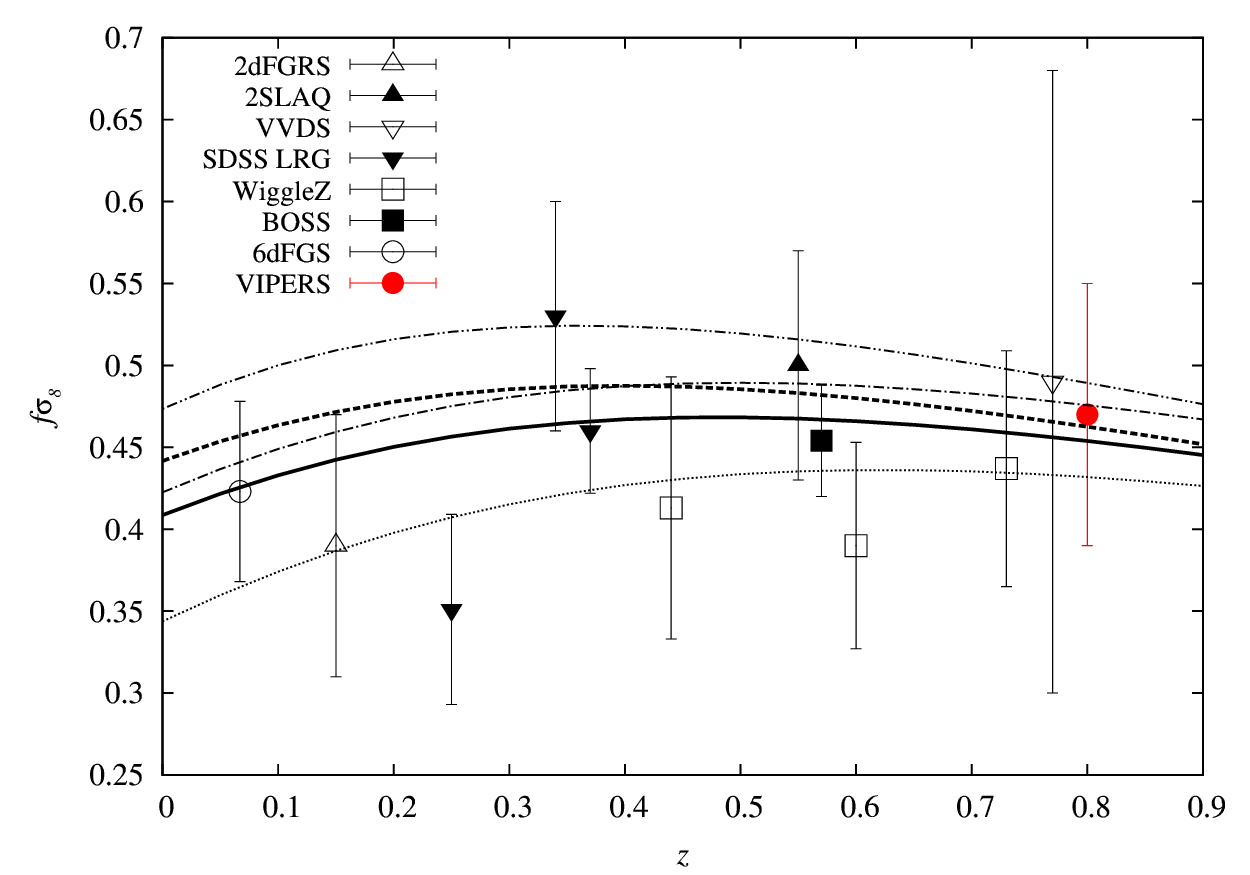}}
\caption{Current measures of  $f(z)\sigma_8$ at different redshifts. The thick solid (dashed) curve shows the General
Relativity prediction in a $\Lambda$CDM model with WMAP9 (Planck) parameters, while the dotted, dot-dashed, and dot-dot-dashed curves show the DGP, coupled dark energy, and $f(R)$ models, respectively. From \cite{delatorre13}.}
\label{fig:fs8}
\end{figure}

In  the next decade or two, ongoing and upcoming wide-field imaging and spectroscopic redshift surveys (e.g., DES, eBOSS, DESI, PFS, LSST, Euclid, WFIRST) aim to measure the growth rate of  cosmic structure on linear and
mildly non-linear scales, up to wave numbers $k\sim 0.2 h$~Mpc$^{-1}$, 
 in the redshift range $0<z<2$. 
A quantity that has been commonly measured  is $f(z)\sigma_8$, where $f(z)=d \ln D/d\ln a$, 
$a=1/(1+z)$ is the scale factor, $D(a)$ the linear growth factor, and $\sigma_8$ 
the normalization of the power spectrum of the density perturbations. The claimed accuracy of the growth rate measured, for example, with Euclid is ~1\% to 2.5\% in the redshift range $0<z<2$ 
\cite{amendola13}; however, 
current measures of $f(z)\sigma_8$ based on redshift space distortions up to redshift $z=0.8$ are 
affected by uncertainties between 10\% and 50\% \cite{howlett14} (Figure \ref{fig:fs8}). Combining  galaxy-galaxy lensing with redshift-space distortions is a promising improvement that, however, is not yet accurate enough to discriminate among different cosmological models and modified gravity theories \cite{reyes10}.

On very non-linear scales, above the wave numbers $k\sim 0.2 h$~Mpc$^{-1}$, the measurement of the growth
rate of individual dark matter halos has only been attempted on the scales
of galaxies: by combining the number of observed pairs of close or disturbed galaxies with the merger probability and
time scale derived from $N$-body simulations, we can infer 
the galaxy-galaxy merger rate \cite{lotz11,jian12,casteels14}. However, current results are inconclusive
\cite{lotz11}: in fact, the merger rate of dark
matter haloes and the merger rate of galaxies are related by dissipative processes that are difficult to model \cite{hopkins13} and these two rates do not necessarily coincide \cite{fakhouri08,guo08,moster13}. 

Dissipative processes are less relevant during the mass accretion of galaxy clusters, whose
rate could be simply estimated based on the 
measurement of the amount of mass in the cluster outskirts. However, this advantage 
over the galaxy-galaxy merger rate has never been capitalized because, in the large and less dense cluster outskirts, (1) cluster galaxies are difficult to distinguish from foreground and background galaxies, and (2) other probes, e.g. X-ray emission, are below the sensitivity of current instruments. Moreover, the cluster outskirts are not in dynamical equilibrium and the usual mass estimation methods based on virial equilibrium are  inappropriate. 

This observational deficiency clashes with the numerous and detailed studies of
the mass growth of galaxy clusters in $N$-body simulations based
on the identification of their merger trees \cite{fakhouri08,vandenbosch02,mcbride09,fakhouri10,giocoli12}. 
 Various laws for the average mass accretion history of dark matter halos
have been proposed:  (1) $M(z)=M_0\exp(-az)$ where $M_0/2$ is the halo mass at the formation redshift of the halo $z_f=\ln 2/a$ \cite{wechsler02}; (2) $\log[M(z)/M_0]=-0.301[\log(1+z)/\log(1+z_f)]^n$ \cite{vandenbosch02}; (3) $M(z)=M_0(1+z)^b\exp(-\gamma z)$ \cite{mcbride09,tasitsiomi04}. The discrepancies between the various relations are due to different mass and time resolutions and different halo statistics of the $N$-body simulations used to infer the relations.  

The Caustic group\footnote{\tt www.dfg.unito.it/ricerca/caustic} in Torino has 
started a project aiming at estimating the mass accretion rate of galaxy clusters
by measuring the mass of a spherical shell surrounding the cluster and its infall time. 
This approach is rather crude when compared with the stochastic aggregation of dark matter halos 
in the hierarchical clustering formation models. Nevertheless, our preliminary
results are promising: they suggest that measuring the mass accretion rate of galaxy clusters
is actually feasible and can potentially provide a new observational test of the cosmological
and structure formation models.

\section{Cluster mass profiles in the outer regions: the caustic method}

If we can measure the mass profile of a cluster beyond its virial radius $r_{200}$, 
we can estimate its instantaneous mass accretion rate as 
\begin{equation}
\dot{M}={M[<r_{200}(1+\delta_s)]-M_{200}\over {\Delta t}} (1+z)^{3/2}  
\label{profileMAR}
\end{equation}
where $M_{200}=M(<r_{200})$.  
Equation (\ref{profileMAR}) assumes that a shell of proper radii $r_{200}$ and $r_{200}(1+\delta_s)$
takes a cosmic time $\Delta t$ to fall onto the cluster with constant acceleration $-GM_{200}/[(1+\delta_s/2)r_{200}]^2$ and null initial velocity. We keep $\Delta t=0.1$~Gyr fixed with redshift and accordingly vary $\delta_s=100 H^2(z) \Delta t^2 $, where $H(z)$ is the Hubble parameter.
The additional factor $(1+z)^{3/2}$ corrects for the transformation from the cosmic time 
$\Delta t$ to the infall time derived with the proper radii $r_{200}$ and $r_{200}(1+\delta_s)$ \cite{deboni15}.

\begin{figure}
 \centerline{\includegraphics[width=0.7\textwidth]{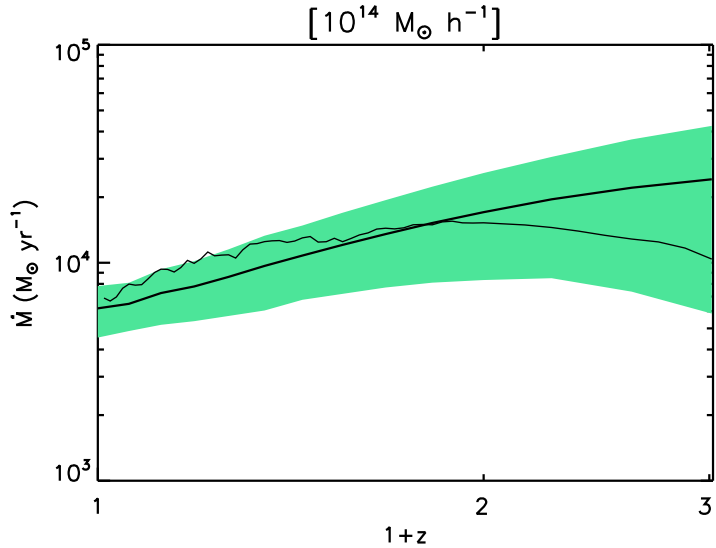}}
\caption{Comparison between our estimate of the mass  accretion rate of a sample of $M_{200}=10^{14}h^{-1}M_\odot$ clusters and $N$-body simulations. The shaded area 
shows the $1\sigma$ range around the mean mass accretion rate estimated with equation (2.1) 
(thick solid line). The thin solid line is the mean mass accretion rate derived from the merger trees. 
Adapted from \cite{deboni15}.} 
\label{fig:mdot}
\end{figure}

Figure \ref{fig:mdot} shows how this simple recipe compares with the CoDECS $N$-body simulation \cite{baldi12} of 
clusters in a $\Lambda$CDM model. 
The thick line shows the mean $\dot{M}$
computed with equation (\ref{profileMAR}) for a sample of clusters with $M_{200}=10^{14}h^{-1} M_\odot$ at $z=0$,
 whereas the thin line
is the mean mass accretion
rate obtained from the merger trees of the dark matter halos. 
The mean accretion rate from the merger trees lie in the region of one standard deviation of our $\dot{M}$ (shaded
area). In addition,  the mean rate from the merger trees is recovered by the mean of our prescription 
within 20\% in the redshift range $z=[0,1]$.

\begin{figure}
 \centerline{\includegraphics[width=0.7\textwidth]{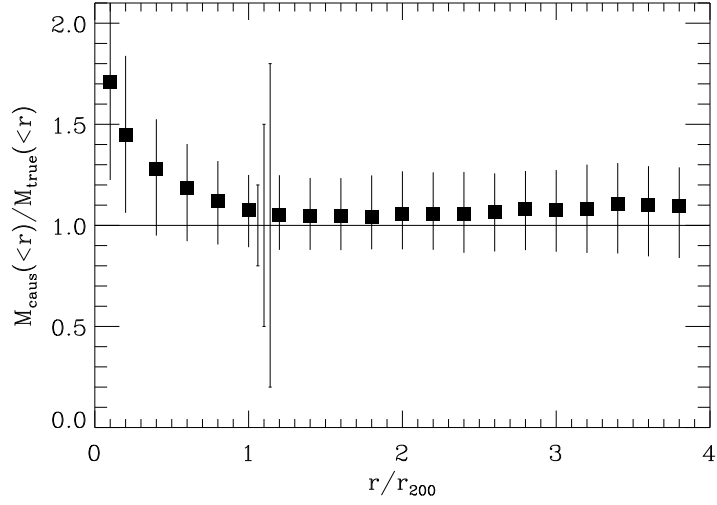}}
\caption{ Mass accuracy with the caustic technique: the figure shows the average ratio between the caustic estimate based on 200 galaxy redshifts and the true mass of simulated clusters. Error bars show the $1\sigma$ uncertainties that mostly originate from projection effects. The overestimate at $r<r_{200}$ is due to a known systematic error \cite{serra11}: at any rate, it is irrelevant in the estimate of the accretion rate with equation (2.1), that involves a mass difference. The three error bars at $r\sim r_{200}$ show the uncertainty on $M_{200}$ estimated with weak lensing for a $10^{15} h^{-1} M_\odot$ cluster at $z=0.3$ (20\%), $z=0.1$ (50\%) and for a $5\times 10^{14} h^{-1} M_\odot$ cluster at $z=0.1$ (80\%) \cite{hoekstra03}. }
\label{fig:mcaus}
\end{figure}

The results of Figure \ref{fig:mdot} show that estimating the mass accretion rate of clusters
is indeed feasible if we can measure the cluster mass profile beyond $r_{200}$. 
This measurement can be performed with the caustic technique. The caustic technique \cite{diaferio97,diaferio99,serra11} uses the celestial coordinates and redshifts of galaxies to estimate the gravitational potential and mass profiles of a cluster from the central region to radii much larger than the virial radius. The caustic technique 
(i) does not rely on the dynamical equilibrium of the cluster; 
(ii) measures the three-dimensional distribution of mass, based on the assumption of spherical symmetry;
(iii) estimates a mass that  is unaffected by the presence of substructures within the cluster \cite{diaferio99,serra11} and by structures along the line of sight \cite{geller13}; 
(iv) can be applied to clusters at any redshift and it is only limited by the telescope time required to measure a sufficiently large number of galaxy redshifts. 
A robust estimate of the cluster mass out to $3r_{200}$ requires $\sim 200$ galaxy redshifts \cite{serra11}. Figure \ref{fig:mcaus} shows how, with this number of redshifts in the field of view of a cluster, the caustic technique recovers the mass profile up to $4r_{200}$ with no bias (solid squares) and with a $1\sigma$ relative uncertainty of 20\% (error bars). The required number of galaxies for such an accurate estimate was demandingly large in the late nineties, when the technique was designed, but, thanks to the development of multi-fiber spectroscopy, it is a feasible target nowadays.

The measurement of a cluster mass accretion rate is tightly linked to the estimate of the total mass of the cluster within its turnaround radius, the so-called ultimate mass $M_u$ \cite{rines06}: with the caustic technique, by combining the 50 CIRS clusters \cite{rines06} with the 58 HeCS clusters \cite{rines13}, we found $M_u/M_{200}=1.99\pm 0.11$, a measure accurate to 5\% \cite{rines13}. Our measure agrees with the $\Lambda$CDM prediction, where $M_u /M_{200}$ has a log-normal distribution with a peak at
mass ratio 2.2 and dispersion 0.38 \cite{busha05}. The accuracy of this unique estimate of the ultimate mass $M_u$ suggests that we can aim to measure the mass accretion rate with the caustic technique to a 20\% accuracy or better. 

\begin{figure}
\centerline{\includegraphics[width=0.7\textwidth]{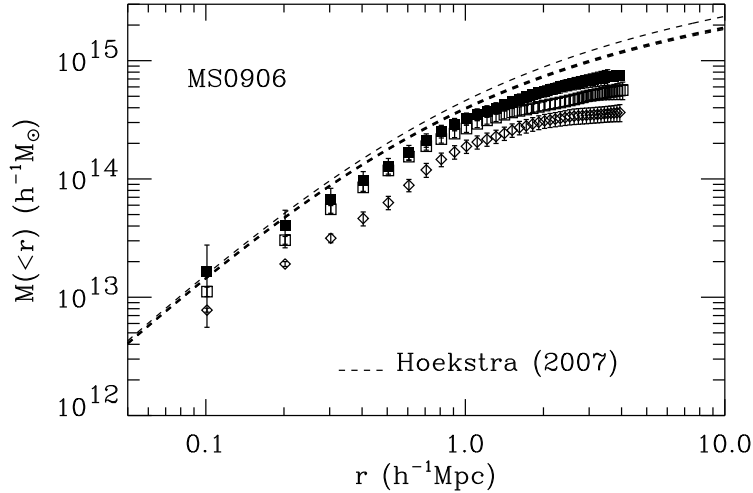}}
\caption{Comparison of the weak lensing mass profile for MS0906 with 1$\sigma$ errors (heavy and light dashed curves respectively, \cite{hoekstra07}) with the mass profiles derived with the caustic technique for MS0906 (open diamonds), A750 (open squares), and  the effective sum of the
caustic mass profiles for the two superposed clusters taking the 0.6$h^{-1}$ Mpc offset between the centers into account (solid squares). From \cite{geller13}.}
\label{fig:mass0906.ps}
\end{figure}

In principle, we could also estimate the mass in the outskirts of clusters with weak gravitational lensing analyses. However, the major source of uncertainty with this approach is the mass projected along the line of sight. This uncertainty increases with decreasing cluster mass and increases with increasing radius. More importantly, the uncertainty increases at both lower and higher redshifts. 
Figure \ref{fig:mcaus} shows that the expected accuracy of the caustic technique is a factor of two better than the accuracy expected with weak lensing for clusters at $z=0.3$, and a factor of four better for clusters at $z=0.1$ \cite{hoekstra03,hoekstra11}. Finally, for the accretion rate estimate, the caustic method can be applied to clusters at any redshift, whereas weak lensing is basically limited to clusters within the redshift range 0.1-0.5 \cite{hoekstra03,hoekstra11}. 

Figure \ref{fig:mass0906.ps} shows a remarkable example of how a superposition along the line of sight 
can affect the weak lensing mass: the two clusters A750 and MS0906 are almost aligned along
the line of sight and the weak lensing mass of MS0906 is approximately the sum of the masses of the two clusters; a sufficiently dense
redshift survey shows that the two clusters are distinct and the caustic technique can easily estimate their 
individual masses \cite{geller13}.

\section{Identification of cluster substructures with the caustic technique}

An additional diagnostic of the mass growth of clusters is 
the presence  of substructures
in the central region and the outskirts of the cluster. 
The first most serious difficulty of this approach is the identification of substructures. Suggested methods are
based on  the distribution of galaxies in redshift space \cite{dressler88}, 
X-ray emission \cite{mohr93,parekh14}, and gravitational lensing \cite{shirasaki14} 
 (see \cite{yu15} for additional references). Some investigations  
have attempted to use this piece of information to constrain the dynamical state of clusters or
their formation time  (see e.g., \cite{mohr96, smith08, deb12, lemze13}). 

With optical observations, the best approach clearly is to combine galaxy positions and redshifts. 
The first step of the caustic method is to arrange the galaxies in a binary
tree based on a galaxy pairwise {\it projected} binding energy, similarly to the method proposed by Serna and Gerbal
\cite{serna96}. However, the caustic method goes further and identifies two thresholds 
that separate the branches of the binary tree and provide a 
list of groups and  a list of substructures from the cluster center out to its outskirts.

The substructures identified in redshift space can correspond to real substructures or can be due to chance 
alignment; in addition, some of the real substructures
of the cluster may not be identified. The number of false detections and the completeness of the substructure sample
depend on the cluster mass, on the substructure mass and, crucially, on the density of the redshift 
survey. When applied to mock redshift surverys of clusters extracted from $N$-body simulations, 
with typical properties of surverys like CIRS \cite{rines06} and HeCS \cite{rines13}, namely
$\sim 200$ redshifts within $3r_{200}$ and cluster mass $M_{200}\sim 10^{14}h^{-1}M_\odot$,
$\sim 50\%$ of the substructures identified with the
caustic technique are false detections, while  the completeness
is between 60\% and 80\% for substructures masses larger than a few $10^{13}h^{-1}M_\odot$ \cite{yu15}.

We thus conclude that the caustic technique appears to be a very promising method to identify substructures
in galaxy clusters. In future work, we will investigate how this technique compares with
methods based on X-ray and gravitational lensing studies. \\ 

I am grateful to my collaborators Ana Laura Serra, Cristiano De Boni, and Heng Yu for 
making this old idea come true and for allowing me to show results from 
our common projects. I thank Christian Marinoni and the other organizers of 
{\it Frontiers of Fundamental Physics 14}  
for giving me the opportunity of illustrating this project at its early stage. 
I am thankful to Margaret Geller and Ken Rines for 
our long-lasting collaboration and 
the countless enlightening discussions on galaxy clusters over the years. 
I acknowledge partial support from the grant Progetti di
Ateneo/CSP TO Call2 2012 0011 "Marco Polo" of the University of Torino, the INFN grant InDark, and the grant
PRIN 2012 "Fisica Astroparticellare Teorica" of the Italian
Ministry of University and Research.

\end{document}